# Biofluorescent Worlds: Global biological fluorescence as a biosignature

Jack T. O'Malley-James[1*] and L. Kaltenegger[1]
[1]*Carl Sagan Institute, Cornell University, Ithaca, NY 14853, USA*
*jomalleyjames@astro.cornell.edu*

**ABSTRACT**
In this paper, we analyze a new possible biological surface feature for habitable worlds orbiting other stars: biofluorescence. High ultraviolet (UV) and blue radiation fluxes drive the strongest biofluorescence in terrestrial fluorescent pigments and proteins. F stars emit more blue and UV radiation than the Sun, while planets and exomoons orbiting such stars remain in the habitable zone for 2-4 Gyr; a timespan that could allow a complex biosphere to develop. Therefore we propose biofluorescence as a new surface biosignature for F star planets. We investigate how the extra emission from surface fluorescence could cause observable signals at specific wavelengths in the visible spectrum. Using the absorption and emission characteristics of common coral fluorescent pigments and proteins, we simulate the increased emission at specific visible wavelengths caused by strong fluorescence, accounting for the effects of different (non-fluorescent) surface features, atmospheric absorption and cloud-cover. Our model shows that exoplanets with a fluorescent biosphere could have characteristic surface colours that allow the presence of surface life to be inferred from observations with upcoming telescopes.

## 1 INTRODUCTION

Fluorescence – the emission of light by a substance that has absorbed light of a shorter wavelength – is widespread in the natural world. A wide range of organisms contain biomolecules that fluoresce, such as chlorophyll in vegetation, green fluorescent proteins (GFPs) and GFP-like proteins in a variety of marine life, or fluorescent compounds/pigments in some land animals, such as insects and amphibians (Holovachov, 2015; Middleton et al., 2015; Sparks et al., 2014; Gruber et al., 2015; Gruber & Sparks, 2015; Taboada et al. 2017). On Earth, fluorescence in vegetation produces a planet-wide effect (shown in Fig.1) that can be detected from orbit (Joiner et al., 2011; Wolanin et al., 2015). Biofluorescence therefore presents previously unexplored possibilities for global spectral biosignatures on extrasolar worlds. In particular, blue wavelengths are known to be effective fluorescence excitation wavelengths; producing the strongest fluorescent responses (Mazel & Fuchs, 2003), making biofluorescence a possible global biosignature for exoplanets orbiting F stars, which emit more of their light at shorter, bluer, wavelengths than the Sun. Here, we explore the idea of a biofluorescent exo-biosphere and its detectability.

Biofluorescent organisms will fluoresce as long as there is a source of radiation that excites fluorescent proteins, or pigments. On Earth, this can produce a detectable global fluorescence signal in the case of chlorophyll fluorescence from surface vegetation (see Fig. 1). However, the extra light emitted by fluorescence in these cases is small compared to the reflected visible light from the planet. Surface vegetation fluorescence causes an approximately 1-2% increase in flux at the fluorescence emission peak wavelength for chlorophyll a. Biofluorescence on Earth is also observed in corals, which can fluoresce with a higher fluorescence efficiency than vegetation. However, biofluorescent corals cover only ~0.2% of the ocean floor, which results in a change in Earth's globally averaged visible flux of just a fraction of a percent. This has a negligible impact on the spectrum of Earth seen as an exoplanet (see results for a more detailed exploration of an Earth-like biofluorescence signal strength). Precise, high resolution observations from Earth-orbiting satellites can disentangle such small signals; however, compared to atmospheric biosignatures or other surface features (see review by Kaltenegger 2017), signals that change the overall planetary flux by less than a percent will not readily be observable for Earth-like planets orbiting other stars. Yet, given the range of exoplanet radiation environments and potential evolutionary histories, biofluorescence could be a potentially detectable spectral surface biosignature for exoplanets hosting widespread biofluorescent biospheres, especially those orbiting stars with larger UV and blue fluxes than the Sun.

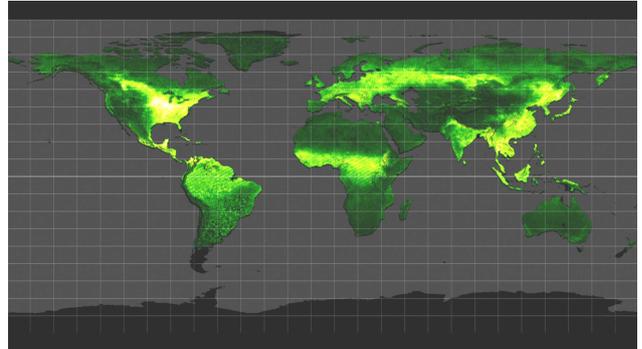

**Figure 1.** Vegetation fluorescence on Earth. Chlorophyll fluorescence can be seen in satellite imagery by disentangling this faint, but unique, signal from other reflectance features, revealing Earth's fluorescent biosignature. The strength and distribution of this signal changes with the seasons and correlates with photosynthetic productivity and vegetation health. Image: *NASA/GSFC.*

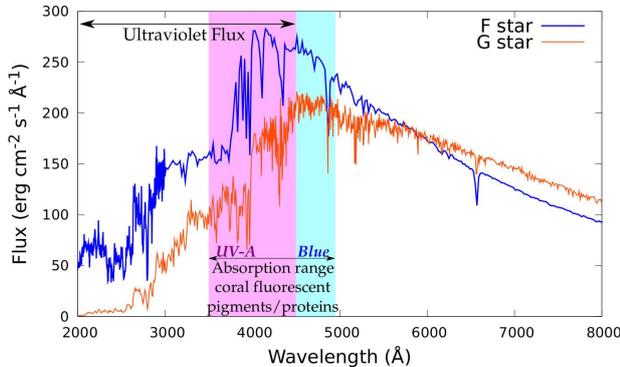

**Figure 2.** Comparing the flux from a sun-like star and an F0 star received by a planet orbiting at a 1 AU equivalent distance. The highlighted regions (UV-A to blue) represent the wavelengths that produce the strongest fluorescence responses in terrestrial corals. A planet orbiting an F-star would receive more flux at these wavelengths than Earth. Note that atmospheric $CO_2$ cuts off the surface UV flux on a planet shortward of 200 nm (see e.g. Rugheimer et al. 2015; O'Malley-James & Kaltenegger 2017).

Therefore, in this paper we explore conditions under which global biofluorescence could be strong enough – i.e. a large surface coverage of biofluorescent life, combined with a higher exciting radiation flux from the host star and high fluorescence efficiency – to cause observable effects in the spectrum of an exoplanet. Given that biofluorescent absorption wavelengths typically fall within the blue and UV part of the spectrum, exoplanet environments with high fluxes of these wavelengths could produce strong biofluorescent signals. Such environments could be found on planets in the habitable zones (HZ) of F-type stars, which have a higher blue and UV radiation output than the Sun (see Fig. 2).

The HZ is defined as the space around a star where liquid water on the surface of a planet is possible (see e.g. Kasting et al., 1993; Kopparapu et al., 2013; Ramirez & Kaltenegger 2017). Hotter, more massive stars than F stars would have an even higher blue/UV output. However, these are rarer and would have very short main sequence lifetimes, making them unsuitable candidates for hosting Earth-like biospheres over billions of years (see, for example, Sato et al., 2014). F stars have HZ lifetimes spanning between 2-4 Gyr (Kasting et al., 1993; Rushby et al., 2013; Sato et al., 2014), a timespan over which complex life could evolve (see e.g. Bounama et al., 2007; Gowanlock et al., 2011).

Planets in the HZs of F stars would be exposed to greater UV fluxes than Earth (see Fig. 2), which would be hazardous for surface biota. When UV radiation is absorbed by biological molecules, especially nucleic acids, harmful effects, such as mutation or inactivation can result, with shorter UV wavelengths having the most damaging effects (see e.g. Voet et al. 1963; Diffey, 1991; Matsunaga et al. 1991; Tevini 1993; Cockell, 1998; Kerwin & Remmele, 2007). Some reef building corals may be using fluorescence as a method of UV-protection. By stepping-up harmful UV wavelengths to longer, safer wavelengths, fluorescence could help corals to prevent oxidative stress, via photochemical reactions, or damage to symbiotic algae (Salih et al., 2000; Takahashi & Murata, 2008; Roth et al., 2010). Hence, we postulate that UV protection in higher lifeforms could be a motivation for the evolution of strong, persistent biofluorescence. This is one of many possible evolutionary trajectories a biosphere could take in high UV environments. However, other UV-protection strategies, such as using UV-absorbing pigmentation, or living in sheltered habitats (underground or underwater) have been considered in previous exoplanet biosignature studies (e.g. Cockell, Kaltenegger & Raven 2009). Biofluorescence as an exoplanet biosignature in such environments has not been evaluated before; hence we focus on this UV-protection strategy here. We explore the increase in visible flux and corresponding change in the spectra of planets in F star systems for a biofluorescent biosphere, for Earth-like biofluorescence based on present-day coral biofluorescence, as well as highly efficient biofluorescence to account for the possibility of the evolution of biofluorescence as a UV defence strategy on such planets. We postulate that terrestrial planets in the HZs of F stars are good targets for surface fluorescent biosignature searches.

## 2 BIOFLUORESCENCE IN CORALS

The process of fluorescence begins when absorbed radiation excites an orbital electron in a fluorescent molecule to its first excited singlet state. Fluorescence emission occurs when this electron relaxes to its ground state. Some of the absorbed energy is released as heat, and some via the emission of a longer wavelength (lower energy) photon than the photon that was originally absorbed; a process known as the Stokes Shift. Fluorescence is near-instantaneous (taking place over nanosecond timescales); however, a related process, phosphorescence involves a different relaxation pathway that results in the delayed emission of light over the course of minutes or hours. The strength of the emitted flux depends on the light environment in the absorption wavelength range and the efficiency of the fluorescence process (or quantum yield), defined as the number of photons emitted to the number absorbed. An efficiency of 1.0 means that each photon absorbed results in an emitted photon, achieving the strongest possible fluorescent flux for a given substance. Many biofluorescent proteins and pigments absorb strongly in the UV and blue parts of the electromagnetic spectrum, emitting visible light photons when they fluoresce. Note that biofluorecsence differs from bioluminescence, which involves exploiting chemical reactions to generate light and is independent of the radiation environment an organism is exposed to.

| Emission peak (nm) | Excitation range (nm) | Fluorescent efficiency (%) on Earth |
|---|---|---|
| 486 | 350-475 | 3-5 |
| 515 | 350-525 | 10-12 |
| 575 | 350-575 | 8-10 |
| 685 | 350-650 | 1-2 |

**Table 1.** The four most common fluorescent pigments and proteins in corals based on coral species selected for being highly fluorescent. Mazel & Fuchs (2003).

On Earth, some of the strongest biological fluorescence results from green fluorescent proteins (GFPs), which are found in numerous marine organisms. Scleractinian (hard) corals are known to fluoresce at a variety of wavelengths, using either GFPs or similar, related proteins (homologs), and fluorescence can have a significant effect on the appearance of coral reefs (Fuchs, 2001). Spectral studies of coral specimens have shown that there are four common pigments/proteins, found either individually or in combination, that account for the majority of observed wavelengths of coral fluorescence (Mazel, 1997). These have approximate fluorescence emission peaks that cover the full range of the visible spectrum at 486 nm (cyan), 515 nm (green), 575 nm (orange) and 685 nm (red). The cyan and green fluorescence is due to GFP-like proteins, the orange fluorescence is caused either by a GFP-like protein or a phycoerythrin from symbiotic cyanobacteria, while the red fluorescence is due to chlorophyll in symbiotic algae (Zawada & Mazel, 2014). The optical signal from a fluorescing coral is composed of reflected light (elastic scatter) and the emitted light from biofluorescence (inelastic scatter). The clustered and static nature of coral reefs provides a template for investigating bright fluorescence as an exoplanet biosignature. Hence we explore whether the additional emitted visible flux due to biofluorescence could become detectable in the overall visible light of a planet.

## 3 METHODS

We explore the remote detectability of global biofluorescence by basing this exo-biosphere on the known spectral properties of terrestrial corals (see Fig. 3). We simulate coral fluorescence by using models of the absorption and emission profiles of the four common fluorescent pigments and proteins in corals, which have fluorescence peaks at 486 nm, 515 nm, 575 nm and 685 nm (see Tab. 1), combined with results from a coupled 1D radiative-convective atmosphere code developed for rocky exoplanets (EXO-Prime; details in Kaltenegger & Sasselov 2010) showing the photon flux reaching the surface of a planet at an Earth-equivalent distance within the HZ of an F0 star. We combine these simulated emission profiles with the coral reflectance spectra (sourced from Roelfsema & Phinn 2006; Clark 2007). An example of the change in the reflectance spectrum caused by simulated fluorescence is illustrated in Fig. 3(iii).

We use the efficiency limits of terrestrial fluorescent proteins as a guide to our exploration of the magnitude of our modelled biofluorescence. The first fluorescent proteins studied were green fluorescent proteins (GFPs), extracted from jellyfish (Shimomura et al., 1962; Johnson et al., 1962; Morin & Hastings, 1971; Monrise et al., 1974; Tsien, 1998). Over time GFPs have been adapted and engineered for use in a variety of applications, from fluorescent microscopy to transgenic pets (see e.g. Stewart (2006) and references therein). GFPs have been engineered in the lab to have a much higher fluorescence efficiency by taking advantage of useful mutations. This has resulted in proteins with efficiencies of up to 100% (Ilagan et al. 2010; Goedhart et al., 2012). Furthermore, highly-efficient fluorescence has been observed in nature, for example, the trees *Pterocarpus indicus* and *Eysenhardtia polystachya* produce matlaline, which fluoresces blue with an efficiency of almost 100% (Lagorio et al. 2015). Therefore, it is feasible that, given the right evolutionary conditions, highly efficient fluorescent pigments or proteins could evolve. For a planet at a 1 AU-equivalent distance around an F0 star, with a present-day Earth-like atmosphere, Rugheimer et al. (2015) modelled a diurnally averaged surface UV-A flux of 52.7 Wm$^{-2}$ (73.6 Wm$^{-2}$ at the top of the atmosphere). The diurnally averaged surface UV-A flux on Earth is 31.5 Wm$^{-2}$ (Rugheimer et al., 2015). Therefore, biofluorescence using GFP-like proteins on exoplanets orbiting F stars could produce a stronger emission signal than on Earth. Furthermore, if fluorescent pigments or proteins evolve to be more efficient on such a planet (as shown to be possible in the lab experiments described above), the additional visible flux could be even higher, potentially producing a remotely observable signal.

Most corals gain energy, carbohydrates and oxygen via symbiotic relationships with algae, which in return feed on $CO_2$ and waste products from the coral. To maintain this symbiosis, coral habitats are limited to the euphotic zone of the oceans (to a maximum depth of 60 m; see e.g. Salih et al., 2000) to allow access to photosynthetically active radiation (PAR). Water attenuation would reduce coral fluorescent flux. Here we make the simplifying assumption of shallow, transparent oceans. This is consistent with shallow-water coral reef habitats on Earth, which are typically warm, clear seawater environments, at depths as shallow as 0-3 m (Kleypas et al., 1999;

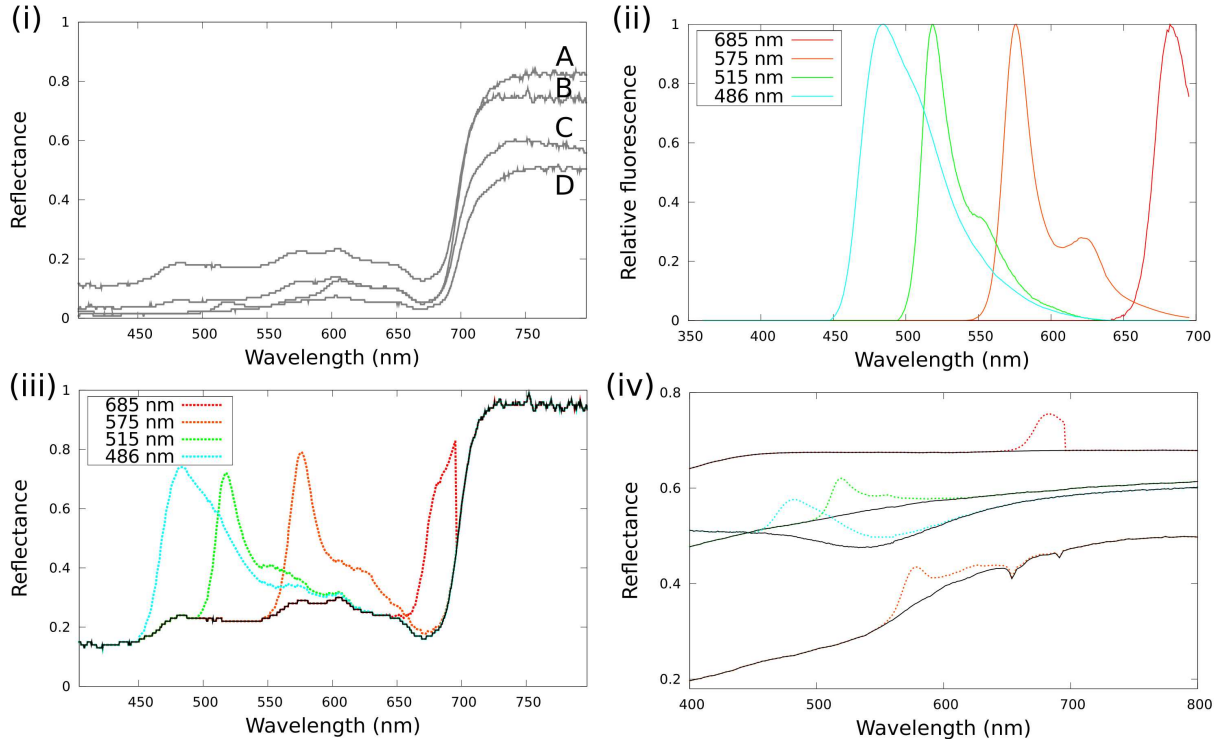

**Figure 3.** **(i)** The reflectance of four different coral species (labelled A, B, C and D). Coral exhibits a "red edge" as a result of chlorophyll in symbiotic algae. A: encrusting coral (low-growing, not branched); B: Acropora spp. (stony, branched); C: Acropora spp. (stony, branched); D: digitate coral (hard, pillar-like appearance). **(ii)** Emission spectra for the four most common fluorescent pigments/proteins in corals. The 685 nm pigment is from chlorophyll-a in symbiotic algae. Reproduced from Fuchs (2001) with data from C. Mazel. **(iii)** An example of simulated fluorescence for each of the fluorescent pigments/proteins. **(iv)** Examples of fluorescent mineral spectra that fluoresce at similar wavelengths to the coral fluorescent pigments/proteins. Sources: USGS Spectral Library, ASTER spectral library, California Institute of Technology.

Nagelkerken et al., 2000). We begin with the assumption that the surface ocean is globally inhabited by biofluorescent life. We then use an ocean spectrum (from the USGS Spectral Library), as an additional surface layer, to explore the effect of different fractions of inhabited versus uninhabited ocean surfaces on the planet's spectrum.

Next we add an atmosphere to the coral surface using EXO-Prime (details in Kaltenegger & Sasselov 2010). The code incorporates a 1D climate (Kasting & Ackerman 1986; Pavlov et al. 2000; HaqqMisra et al. 2008), 1D photochemistry (Pavlov & Kasting 2002; Segura et al. 2005, 2007), and 1D radiative transfer model (Traub & Stier 1976; Kaltenegger & Traub 2009) that can be used to calculate the model spectrum of an Earth-like exoplanet orbiting different host stars in the HZ. EXO-Prime is a model that simulates both the effects of stellar radiation on a planetary environment and the planet's outgoing spectrum.

To simulate how much UV radiation would be available to excite biofluorescence, we first model the surface radiation environment for an exoplanet in the habitable zone of an F star. F stars have higher UV fluxes than G stars, like our own Sun. We then model the generated emitted flux due to biofluorescence for these UV light levels impinging the four pigments/proteins. Our model takes the photon flux over the excitation range for a given fluorescent protein, then uses a range of efficiencies for biofluorescence (up to 1.0, which represents a fluorescent efficiency of 100%) to calculate the additional emitted visible photons that could be generated by a biofluorescence biosphere. This is added to the reflected photon flux at visible wavelengths from our model surfaces to determine the modelled biofluorescent coral spectra that form part of our model planet spectra. Note that our non-excited coral spectra are natural coral spectra that will contain very low fluorescence excitation and emission features caused by Earth's UV radiation environment. However, compared to the UV induced fluorescence signals we model here, the strengths of these features are negligible; hence we add our modelled fluorescence onto the natural spectra, treating them as if they contain no fluorescence features. Initially, we explore a cloud-free case. Then, the effect of clouds is modeled by using a cloud albedo spectrum, which we vary to model different cloud coverages up 50% (an Earth-like cloud fraction); see Kaltenegger et al. (2007) for details.

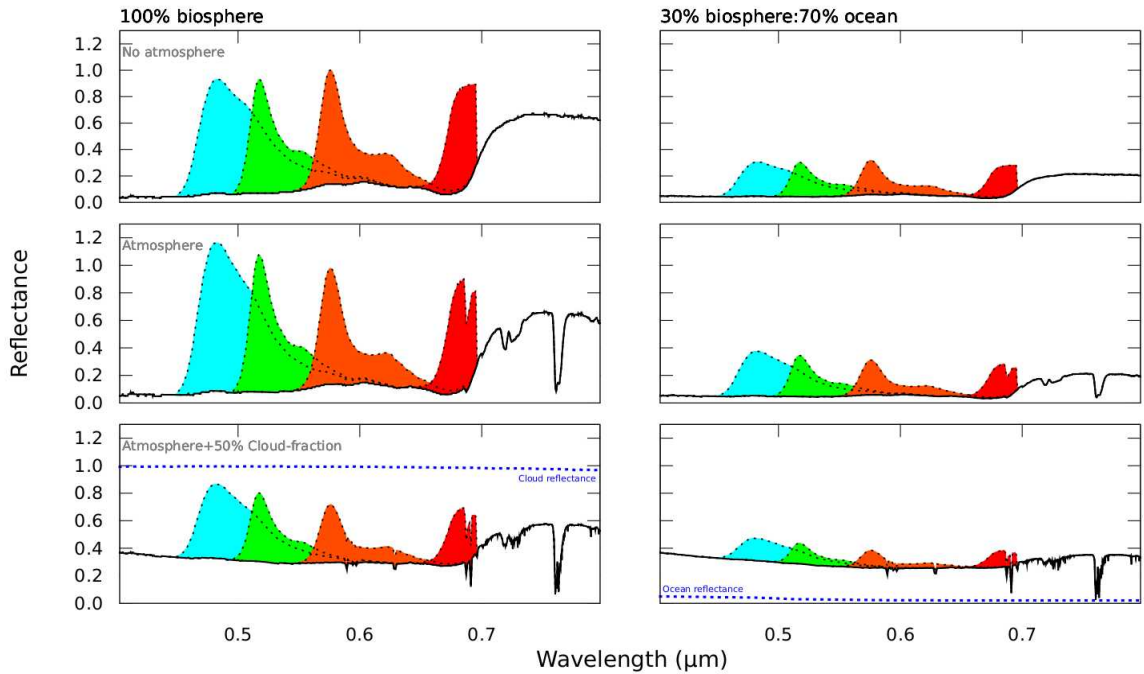

**Figure 4.** The change in the shape of the coral surface spectra (using coral C as an example) for each fluorescent wavelength for (left)100% surface coverage and (right) 30% surface coverage with 70% uninhabited ocean surface. The top panels show the atmosphere-freecase. The middle panels show the effect of adding an atmosphere with clear skies. The bottom panels show an atmosphere with anEarth-like cloud cover of 50%. The individual reflectance profiles (ocean and clouds) used to generate our model surfaces are shown byblue dashed lines in the bottom panels.

**3.1. False-positives**

It is also possible for fluorescence to occur abiotically. Some minerals (e.g. calcite, fluorite, opal, zircon) and polycyclic aromatic hydrocarbons (PAHs; e.g. fluoroanthene, perylene, pyrene) are fluorescent at similar wavelengths to those of fluorescent corals; a result of metal cation impurities in the case of minerals, and delocalised electrons in aromatic molecule groups for the case of PAHs (see e.g. McDougall, 1952; Beltrán et al., 1998). Therefore, it is possible that surface fluorescence could be observed, if the surface of a planet is composed solely of a certain mineral, for example.

However, hydrocarbons and the metal inclusions within fluorescent minerals would not be subject to Darwinian evolution, so it would be unlikely for these to fluoresce strongly enough to be observable. As an additional comparison, we model surfaces based on spectra of fluorescent minerals that fluoresce strongly at similar wavelengths to the coral pigments/proteins (see Fig. 3(iv) versus Fig. 3(iii)).

Future exoplanet characterization observations will be limited by available observing time and resources. Therefore the best candidates will need to be selected from lists of possible target planets. Colour-colour diagrams are a useful tool for prioritizing rocky exoplanets for further follow-up observations. The colours of biological surface features tend to fall within a certain region of a colour-colour diagram that is distinct from areas occupied by ice giants, gas giants, and rocky planets with abiotic surfaces (Hegde et al. 2013; 2015). We used a range of reflectance spectra from different coral species to produce a colour-colour diagram, using the following relation for comparing the reflectance, r, in two different colour bands:

$$C_{AB} = A - B = -2.5\log_{10}(r_A/r_B) \quad (1)$$

Where $C_{AB}$ is the difference between two arbitrary colour bands, A and B. Here we use standard Johnson-Cousins BVI broadband filters to define the colour bands (0.4 μm < B < 0.5 μm; 0.5 μm < V < 0.7 μm; 0.7 μm < I < 0.9 μm).

**4 RESULTS & DISCUSSION**

Biofluorescence emission at 486nm, 515 nm, 575 nm and 685 nm causes notable changes to a planet's surface spectrum (Fig.4), especially in cases where biofluorescent life covers a large fraction of the surface area of a planet and fluoresces with a high efficiency. For the case of biofluorescence on Earth, where coral reefs cover approximately 0.2% of the ocean floor, the signal would result in a fraction of percent change in emitted flux at the peak emission wavelengths, when the flux is averaged over the whole surface of the planet, as it would be for an exoplanet observation. Such a small change would be too weak to detect for upcoming observations with large telescopes like the European Extremely Large Telescope. However, if such a biosphere covered a larger fraction of a planet's surface (30% biofluorescent surface:70% ocean), the globally averaged additional emitted visible flux caused by such biofluorescence under an F0 star radiation environment can result in an increase of the flux of about 10% at peak emission wavelengths. If the surface coverage of such a biofluorescent biosphere is increased further, it would reach biofluorescent emission signal strengths of the order of ~30% at peak emission wavelengths (Table 2). With a cloud-free atmosphere, these signal strengths would be further increased (see Table 2).

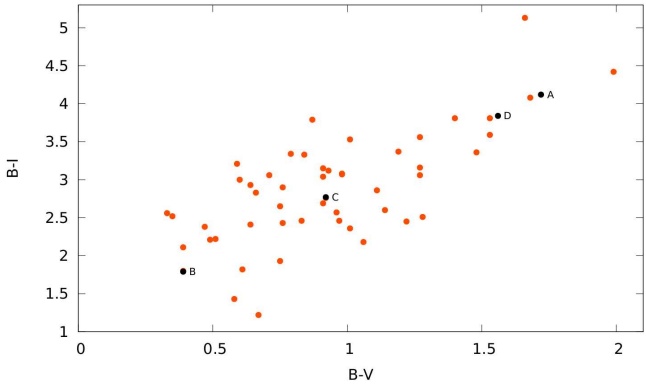

**Figure 5.** A colour-colour diagram showing the wide distribution of a variety of coral species in colour space. The labelled points show the positions the corals we chose for our surface biosphere models where A: encrusting coral (low-growing, not branched); B: Acropora spp. (stony, branched); C: Acropora spp. (stony, branched); D: digitate coral (hard, pillar-like appearance). Source: Roelfsema C. & Phinn S. (2006) "Spectral reflectance library of selected biotic and abiotic coral reef features in Heron Reef."; Clark (2007), Torres-Pérez (private communication).

| Wavelength (nm) | % Change in Visible Flux Caused By Fluorescence | | | |
|---|---|---|---|---|
| | Clear Atmosphere | | Atmosphere+50% cloud-fraction | |
| | *100% surface cover* | *30% surface cover : 70% ocean* | *100% surface cover* | *30% surface cover : 70% ocean* |
| 486 | 40 – 68 | 22 – 37 | 8 – 14 | 5 – 8 |
| 515 | 136 –160 | 75 – 90 | 26 – 31 | 15 – 18 |
| 575 | 56 – 70 | 42 – 53 | 16 – 20 | 12 – 15 |
| 685 | 8 – 16 | 6 – 13 | 2 – 4 | 1 – 3 |

**Table 2.** The change in the outgoing visible flux with and without fluorescence at the peak emission wavelengths of the fluorescent pigments/proteins, assuming Earth-like fluorescent efficiencies (using values ranges from Table 1) under an F0 radiation environment. Values given for full and fractional biofluorescent surface coverage, with and without clouds.

| Wavelength (nm) | % Change in Visible Flux Caused By Fluorescence | | | |
|---|---|---|---|---|
| | Clear Atmosphere | | Atmosphere+50% cloud-fraction | |
| | *100% surface cover* | *30% surface cover : 70% ocean* | *100% surface cover* | *30% surface cover : 70% ocean* |
| 486 | 1350 | 740 | 270 | 150 |
| 515 | 1360 | 750 | 260 | 150 |
| 575 | 700 | 530 | 200 | 150 |
| 685 | 800 | 630 | 200 | 140 |

**Table 3.** The change in the outgoing visible flux with and without fluorescence at the peak emission wavelengths of the fluorescent pigments/proteins, assuming fluorescent efficiencies of 100% under an F0 radiation environment. Values given for full and fractional biofluorescent surface coverage, with and without clouds.

For highly efficient biofluorescence, (Fig. 4 (left column, 100% surface coverage)), the large flux of emitted photons caused by biofluorescence results in a change in visible flux that can be up to 1350% higher than it would be without biofluorescence at the peak fluorescent wavelengths (see Table 3). Note that the combination of Rayleigh scattering in the atmosphere, which is strongest at bluer wavelengths, and biofluorescence emission can result in an apparent reflectance that is greater than 100% at blue wavelengths, because fluorescence introduces additional visible photons that are not part of the reflected component of visible light. For the case where cloud cover is included, the visible flux can still be up to ~270% higher than it would be without biofluorescence at the peak fluorescent wavelengths.

The addition of open ocean without a biofluorescent biosphere (Fig. 4 (right column)) reduces the globally averaged effect of the biofluorescence signal, because water has a very low albedo; however, even for the case where 70% of the surface is covered by uninhabited ocean with 50% cloud coverage, the visible model spectrum is ~150% brighter at peak wavelengths than it would be without biofluorescence.

Biofluorescence also causes a planet's surface colour to occupy a separate region of a colour-colour diagram compared to other surface features such as vegetation and fluorescent minerals. Our model coral spectra span a wide colour space, shown in Fig. 5, allowing us to investigate how these relationships hold for a wide range of coral surfaces. Model fluorescent biospheres are distinguishable from other features, except for the 486 nm model, which occupies a similar region of colour space to vegetation. This shows that in some cases a coral-like surface could be mistaken for vegetation, or vice versa for very low resolution spectra (see Fig. 4). Fluorescent mineral surfaces remain distinct from the corals. This is a general trend observed across the different ocean and cloud fraction scenarios for a planet with an Earth-like atmosphere (Fig. 6 and Fig. 7).

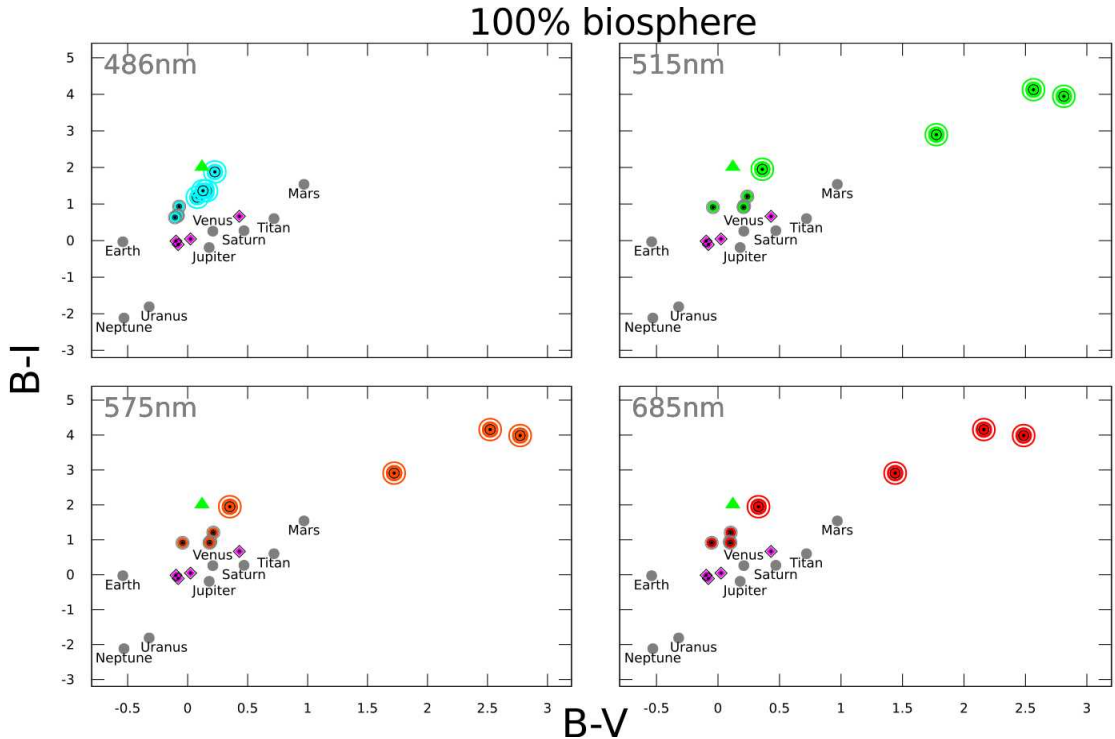

**Figure 6**. Colour-colour diagrams for full surface biosphere coverage for each of the four sample corals. The filled circles show thepositions of fluorescing coral surfaces with an atmosphere and clear skies; the colour indicating the emission colour of the fluorescentpigments/proteins. Grey-ringed circles show the colours for the 50% cloud-cover cases. These are compared to abiotic (mineral) fluorescentsurfaces (diamonds), vegetation (green triangles) and solar system bodies (grey filled circles).

The change in a planet's colour caused by increasing the fraction of uninhabited ocean is shown in Figs. 6 and 7. Note that we derive these colours from theoretical spectra and thus do not add error bars to our calculations. These colour-color diagrams are intended to provide input for instrument simulators for upcoming instruments built to detect the colours of rocky planets. These will have individual errors associated with them, depending on the telescope and instrument used. Increasing the ocean fraction decreases the colour separation between fluorescing corals and abiotic surfaces, but these remain distinct. Increasing cloud cover decreases the separation between fluorescent corals and abiotic surfaces in colour space. However, for an Earth-like cloud cover of 50%, some fluorescent cases at 515 nm, 575 nm and 685 nm remain distinct.

The biofluorescent planet colours remain distinct from the solar system bodies (with the exception of one of the 515 nm fluorescent surfaces with a 70% ocean fraction, which occupies a similar position to Mars; Fig. 7, top right), strengthening the case for colour-colour diagrams to prioritize targets for time-intense follow up spectral observations.

### 4.1. Prospects for detectablity

Terrestrial remote sensing observations can detect the faint fluorescent signal caused by surface vegetation (Joiner et al., 2011; Wolanin et al., 2015), as illustrated in Fig. 1. It is also possible to detect fluorescent corals from orbit. In Lubin et al. (2001), top-of-atmosphere reflectance spectra for corals were compared to sand and algae. They found that there was a large enough contrast difference between corals, sand and coralline algae for corals to be identified using remote sensing techniques. The strong biofluorescence we postulate here could cause an reflectance/emittane feature that exceeds 100% reflectivity, improving the prospects for observing a biofluorescence signature on nearby exoplanets. The spectroscopic appearance of planets with biofluorescent surface life provides additional reference points in the colour space of life (see Hegde et al. 2013), which will help to classify the best targets for life-detection follow-up observations.

The angular resolution needed to directly detect a HZ planet depends on the apparent angular separation of the planet from its host star as viewed from Earth. Planets orbiting in the HZs of F stars have larger angular separations. We summarise the properties of the six closest F stars in Table 4 to illustrate the range of HZ widths F type stars could have, as well as the angular separations of planets at the HZ boundaries. Note that none of these specific examples have known planets in their HZs yet. Upcoming telescopes like the 38 m European Extremely Large Telescope (E-ELT) will have inner working angles as low as 6 milliarcseconds (mas) at visible wavelengths. Assuming a HZ outer edge of 2 AU, an inner working angle of 6 mas will enable us to directly observe planets in the HZs of F stars out to distances of ~300 pc. F type stars make up ~3% of stars in the Milky Way, and there are over 300 F type stars within just 30 pc (100 ly), providing a large set of potential targets.

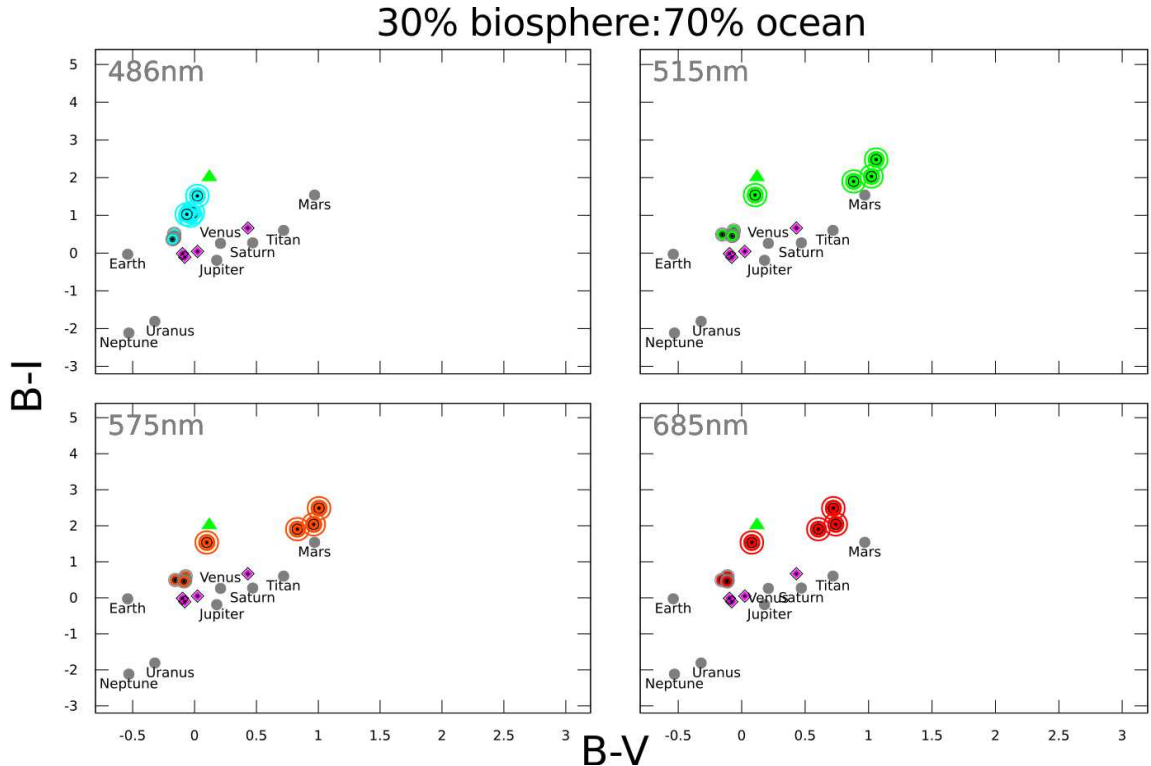

**Figure 7.** Colour-colour diagrams for 30% surface biosphere coverage with 70% open ocean for each of the four sample corals. The filledcircles show the positions of fluorescing coral surfaces with an atmosphere and clear skies; the colour indicating the emission colour ofthe fluorescent pigments/proteins. Grey-ringed circles show the colours for the 50% cloud-cover cases. These are compared to abiotic(mineral) fluorescent surfaces (diamonds), vegetation (green triangles) and solar system bodies (grey filled circles).

For planets with non-complete cloud coverage, the effect of clouds can be distinguished from surface features with many short, high signal-to-noise observations, because clouds should occupy all areas of the planet given enough time. Thus one can separate them from surface features that are bound to the rotation of a planet, if the observations can be limited to about 1/20 of the planet's rotation period, or for the Earth, about an hour (see Pallé et al. 2008).

Moving closer to the inner edge of the HZ increases the UV flux a planet would receive. Hence, a fluorescence signal could become more pronounced for planets orbiting closer to the inner-edge of the HZ. However, Earth-like planets near the inner-edge are expected to have water-rich atmospheres, or to be in the process of losing their atmospheres to space (see e.g. Kasting et al., 1993; Goldblatt & Watson, 2012; Kopparapu et al., 2013), which could increase cloud-cover to such an extent that any surface signal is undetectable. For dry, desert worlds, with limited surface water, the HZ inner edge could have extended to the orbit of Venus in the solar system until as recently as 1 Gyr ago (Abe et al., 2011). Here the UV flux at the top of the atmosphere would be 190% of Earth's TOA flux (Cockell, 1999). Clear skies on a dry planet could aid detection, but the non-Earth-like history of such a planet leaves many questions open about the course the evolution of a biosphere might take. As mentioned earlier, eroded thinner atmospheres would increase surface UV flux and detectable fluorescence features.

For ocean worlds, tidal forces should influence nutrient cycling in ocean waters and hence will play a prominent role in determining the type of biosphere that develops. Strong tidal forces on a planet, such as those caused by a large moon, could lead to periodic nutrient abundances (Lingam & Loeb 2017) that induce population booms in a shallow water biosphere like the one we hypothesise here. This could periodically strengthen a fluorescence signal by increasing the inhabited surface fraction of the planet.

| Star | Type | Dist. (pc) | $L_*$ ($L_\odot$) | $T_{eff}$ (K) | Age[1] (Gyr) | HZ (AU) | Angular Sep. (arcsec) | Volcanic Hydrogen HZ (outer edge; AU) | Angular Sep. (arcsec) |
|---|---|---|---|---|---|---|---|---|---|
| Procyon A | F5 IV-V | 3.53 | 6.9 | 6530 | 1.7 | 1.9-4.4 | 0.54-1.24 | 6.1 | 1.72 |
| $\pi^3$ Orionis | F6 V | 8.06 | 2.8 | 6519 | 1.4 | 1.2-2.8 | 0.15-0.35 | 3.9 | 0.48 |
| $\chi$ Draconis | F7 V | 8.06 | 1.86 | 6150 | 5.3 | 1.0-2.3 | 0.12-0.29 | 3.2 | 0.40 |
| $\zeta$ Tucanae | F9 V | 8.58 | 1.26 | 5970 | 4.0 | 0.8-1.9 | 0.09-0.22 | 2.7 | 0.31 |
| $\gamma$ Leporis A | F7 V | 8.98 | 2.6 | 6299 | 1.3 | 1.2-2.7 | 0.13-0.30 | 3.8 | 0.42 |
| $\gamma$ Pavonis | F6 V | 9.23 | 1.52 | 6112 | 1.1 | 0.9-2.1 | 0.10-0.23 | 2.9 | 0.31 |

**Table 4.** F stars within 10 pc. The HZ was estimated using the methods of Kopparapu et al. (2013), assuming an inner-edge defined by a recent habitable Venus, and an outer-edge by an early habitable Mars. We also show how the volcanic hydrogen HZ could extend the outer edge of the HZ in these systems (see Ramirez & Kaltenegger (2017) for details). Angular separations of HZ planets are calculated as a guide to observability as the ability of a telescope to resolve a planet is determined by its inner working angle; the minimum observable apparent separation of a star and planet as viewed from Earth. Data: RECONS catalog; (1) Age estimates from the Geneva-Copenhagen Survey of Solar Neighbourhood III (Holmberg et al. 2009).

### 4.2. UV surface environments on planets in F star systems

On the present-day Earth, the ozone layer prevents the most damaging UV wavelengths, UV-C radiation (100 – 290 nm), from reaching the surface. However, on other HZ planets, a protective ozone layer may not be present; the early Earth, for example, lacked a significant ozone layer, which could be detectable in its spectra (see e.g. Zahnle et al. 2007; Kaltenegger et al. 2007). Depending on the atmospheric composition of a planet, other atmospheric gases, such as sulphur compounds and $CO_2$, or photochemical hazes can absorb UV radiation (see e.g., Cockell et al., 2000; Rugheimer et al., 2015; Arney et al. 2016; O'Malley James & Kaltenegger 2017). The thinner the atmosphere of a planet is, the more damaging radiation reaches the planet's surface. Therefore mechanisms that protect biota from this radiation are a crucial part of maintaining surface habitability, especially on planets with thin atmospheres that, for example, are less massive and therefore cannot maintain a dense, protective atmosphere, or have had their atmosphere stripped away. Hence, we postulate that UV protection could be a motivation for the evolution of strong, persistent biofluorescence. This is one of many possible evolutionary trajectories a biosphere could take in high UV environments. However, other UV-protection strategies, such as using UV-absorbing pigmentation, or living in sheltered habitats (underground or underwater) have been considered in previous exoplanet biosignature studies (e.g. Cockell, Kaltenegger & Raven 2009). Biofluorescence as exoplanet biosignature in such environments has not been evaluated before; hence we focus on this UV-protection strategy here.

Planets in the HZs of F stars would be exposed to greater UV fluxes than Earth (see Fig. 2), which would be hazardous for surface biota. Furthermore, F stars have shorter HZ lifetimes (2-4 Gyr) than G stars like the Sun. If life follows a similar evolutionary trajectory to life on Earth, F star planets may not develop an oxygen-rich atmosphere until late into their habitable lifetimes (if at all) – on Earth, oxygen didn't reach significant levels in the atmosphere for more than 2 Gyr after the origin of life (see e.g. Zahnle et al. 2013) and did not become detectable until about 3 billion years into Earth's history (see e.g. Kaltenegger et al. 2007). Low ozone levels associated with a low-oxygen atmosphere would lead to high surface UV fluxes for F star planets, which would make the evolution of biofluorescence as a UV protection method favourable for surface life.

### 4.3. Mitigating the biological cost of intense fluorescence

Organisms using FPs under high-intensity radiation regimes would need to overcome the problem of photobleaching. All FPs eventually enter a non-fluorescent state (bleaching) after extended excitation – this is often observed when FPs are used in a lab setting, but also occurs naturally when corals are exposed to strong light sources (Shaner et al., 2005). This is commonly an irreversible reaction, which, given that Fps are expensive molecules to make (see e.g. Eyal et al., 2015), would make bright, continuous biofluorescence biologically costly.

However, some FPs can regain their fluorescent ability after bleaching (Ando et al., 2004; Sinnecker et al., 2005; Henderson et al., 2007; Zhou & Lin, 2013). Reversibly Switchable Fluorescent Proteins (RSFPs) initially respond in the "classical" way to intense excitation, with an exponentially decaying emission intensity. Then, instead of decaying to a non-fluorescent state, the emission from these proteins stabilises at a lower level (Sinnecker et al., 2005). Furthermore, after being in darkness, or exposed to certain non-excitation wavelengths, the proteins regain their original (or close to original) fluorescent intensities (Sinnecker et al., 2005; Henderson et al., 2007). Naturally occurring Fps from corals in the family *Pectiniidae* were further enhanced in the lab via rational mutagenesis and directed evolution to produce bright, photoreversible FPs (Ando et al., 2004; Henerson et al., 2007). Similar forms of FPs could buffer a fluorescent biosphere against photobleaching.

Another solution could be found by avoiding FPs altogether. Quantum dots (inorganic nanocrystals with unique optical and chemical properties) are used as alternative fluorescent markers in some experiments in order to avoid the problems associated with bleaching (see e.g. Resch-Genger et al. 2008). Although these are inorganic, it is not unprecedented for nano-particle sized organic molecules to occur in nature (magnetotactic bacteria use magnetic nanoparticles to align themselves to field lines; Xie et al., 2009); hence, organic nanoparticles that make use of similar physics to fluorescent quantum dots, could result from an alternative course of evolution on another world.

# 5 CONCLUSIONS

In this work, we propose biofluorescence as a new biosignature. We find that a biofluorescent biosphere, based on commonly occurring coral fluorescent pigments/proteins, can alter the surface reflectance spectrum of a planet in a distinct way, which could be used to infer the presence of surface life. We use colour-colour diagrams to illustrate the precision needed to identify a biofluorescent surface biosignature, modelled using the fluorescent efficiencies of the brightest biofluorescent life on Earth. In colour space, these surfaces are distinguishable from abiotic Solar System planets and fluorescent mineral surfaces, because bright fluorescence would have to evolve via Darwinian selection to suit a planet's environmental conditions; whereas abiotic fluorescence would not be subject to such selection. A continuous fluorescence signature could be especially suited to habitable F star planets, which have high UV and blue radiation fluxes. The angular separation of the HZs of F star systems up to 300 pc away will allow telescopes like the E-ELT to explore any rocky planets in the HZ for such surface features directly. While these will be challenging observations to make, theoretical work such as this provides targets to help shape the design of these future instruments to give us the best chance of succesfully making biosignature observations. Life on other worlds could be very different to life on Earth, but speculations about the detectability of extrasolar life are, by necessity, bound by what we know about present, or past, terrestrial life. However, by taking aspects of the terrestrial biosphere and placing them in non-terrestrial environments, it is possible to move towards hypothetical biospheres that are alien, but still rooted in biology.

**ACKNOWLEDGEMENTS**

The authors would like to thank the anonymous reviewer for constructive comments that helped to improve the manuscript. JTO acknowledges helpful discussion with Charles Mazel. Juan Torres-Pérez for providing coral spectral data. Sarah Rugheimer for providing stellar spectra. We also acknowledge funding from the Simons Foundation (290357, Kaltenegger).

**REFERENCES**

Abe Y., Abe-Ouchi A., Sleep N. H., Zahnle K. J., 2011, Astrobiology, 11, 443

Ando R., Mizuno H., Miyawaki A., 2004, Science, 306, 1370

Arney, G. et al. 2016, Astrobiology 16, 73

Beltrán J. L., Ferrer R., Guiteras J., 1998, Analytica chimica acta, 373, 311

Bounama C., von Bloh W., Franck S., 2007, Astrobiology, 7, 745

Clark R. N., 2007, USGS Digital Spectral Library splib06a, Data Series 231

Cockell C. S., 1998, J. Theor Biol, 193, 717

Cockell C. S., 1999, Planetary and Space Science, 47, 1487

Cockell C. S., Catling D. C., Davis W. L., Snook K., Kepner R. L., Lee P., McKay C. P., 2000, Icarus, 146, 343

Cockell C. S., Kaltenegger L., Raven J. A., 2009, Astrobiology, 9, 623

Diffey B. L., 1991, Physics in medicine and biology, 36, 299

Eyal G, et al., 2015, PloS one, 10, e0128697

Fuchs E., 2001, Appl Opt, 40, 3614

Goedhart J., et al., 2012, Nat Commun 3, 751

Goldblatt C., Watson A. J., 2012, Phil Tran R Soc A, 370, 4197

Gowanlock M. G., Patton D. R., McConnell S. M., 2011, Astrobiology, 11, 855

Gruber D. F., Gaffney J. P., Mehr S., DeSalle R., Sparks J. S., Platisa J., Pieribone V. A., 2015, PloS one, 10, e0140972

Gruber D. F., Sparks J. S., 2015, American Museum Novitates, 3845, 1

Haqq-Misra J. D., Domagal-Goldman S. D., Kasting P. J., Kasting J. F., 2008, Astrobiology, 8, 1127

Hegde S., Kaltenegger L., 2013, Astrobiology, 13, 47

Hegde S., Paulino-Lima I. G., Kent R., Kaltenegger L., Rothschild L., 2015, PNAS, 112, 3886

Henderson J. N., Ai H. W., Campbell R. E., Remington S. J., 2007, PNAS, 104, 6672

Holmberg J., Nordstroem B., Andersen J., 2009, Astron Astrophys, 501, 941

Holovachov O., 2015, Green Letters: Studies in Ecocriticism, 19, 329

Ilagan R. P., Rhoades E., Gruber D. F., Kao H. T., Pieribone V. A., Regan L., 2010, FEBS Journal, 277, 1967 Johnson F. H., Shimomura O., Saiga Y., Gershman L. C., Reynolds G. T., Waters J. R., 1962, Journal of Cellular and Comparative Physiology, 60, 85

Joiner J., Yoshida Y., Vasilkov A. P., Middleton E. M., 2011, Biogeosciences, 8, 637

Kaltenegger L., Traub W. A., Jucks K. W., 2007, Astrophys J, 658, 598

Kaltenegger L., Traub W. A., Astrophys J, 698, 519

Kaltenegger L., Sasselov D., 2010, Astrophys J, 708, 1162

Kasting J. F., Ackerman T. P., 1986, Science, 234, 1383

Kasting J. F., Whitmire D. P., Reynolds R. T., 1993, Icarus, 101, 108

Kerwin B. A., Remmele R. L., 2007, Journal of Pharmaceutical Sciences, 96, 1468

Kleypas J. A., McManus J. W., Meñez L. A. B., 1999, Amer Zool, 39, 146

Kopparapu R. K., et al., 2013, Astrophys J, 765, 131

Lagorio M. G., Cordon G. B., Iriel A., 2015, Photochem Photobiol Sci, 14, 1538

Lingam M., Loeb A., 2017, arXiv preprint, arXiv:1707.04594

Lubin D., Li W., Dustan P., Mazel C. H., Stamnes K., 2001, Remote Sens Environ, 75, 127

Matsunaga T., Hieda K., Nikaido O., 1991, Photochemistry and photobiology, 54, 403

Mazel C. H., 1997, Proceedings of SPIE 2963, 240

Mazel C. H., Fuchs E., 2003, Limnol Oceanog, 48, 390

McDougall D. J., 1952, Am Mineral, 37, 427

Middleton E. M., et al., 2015, Geoscience and Remote Sensing Symposium (IGARSS), IEEE, 3878

Morin J. G., Hastings J. W., 1971, Journal of cellular physiology, 77, 313

Morise H., Shimomura O., Johnson F. H., Winant J., 1974, Biochemistry, 13, 2656

Nagelkerken I., Van der Velde G., Gorissen M. W., Meijer G. J., Van't Hof T., Den Hartog C., 2000, Estuar Coast Shelf Sci, 51, 31

O'Malley-James J. T., Kaltenegger L., 2017, MNRAS Lett, 469, L26

Pallé E., Ford E. B., Seager S., Montañés-Rodríguez P., Vazquez M., 2008, Astrophys J, 676, 1319

Pavlov A. A., Kasting J. F., Brown L. L., Rages K. A., Freedman R., 2000, J Geophys Res, 105, 981

Pavlov A. A., Kasting J. F., 2002, Astrobiology, 2, 27

Ramirez R., Kaltenegger L., 2017, Astrophys J Lett, 837, L4

Resch-Genger U., Grabolle M., Cavaliere-Jaricot S., Nitschke R., Nann T., 2008, Nature methods, 5, 763

Roelfsema C., Phinn S., 2006, Spectral reflectance library of selected biotic and abiotic coral reef features In Heron Reef, Bremerhaven, PANGAEA, hdl:10013/epic.40554

Roth M. S., Latz M. I., Goericke R., Deheyn D. D., 2010, J Exp Biology, 213, 3644

Rugheimer S., Segura A., Kaltenegger L., Sasselov D., 2015, Astrophys J, 806, 137

Rushby A. J., Claire M. W., Osborn H., Watson A. J., 2013, Astrobiology, 13, 833

Salih A., Larkum A., Cox G., Kuhl M., Hoegh-Guldberg O., 2000, Nature, 408, 850

Sato S., Cuntz M., Olvera C. G., Jack D., Schroder K. P., 2014, Int J Astrobiol, 13, 244

Segura A., Kasting J. F., Meadows V., Cohen M., Scalo J., Crisp D., Butler R. A., Tinetti G., 2005, Astrobiology, 5, 706

Segura A., Meadows V. S., Kasting J. F., Crisp D., Cohen M., 2007, Astron Astrophys, 472, 665

Shaner N. C., Steinbach P. A., Tsien R. Y., 2005, Nature methods, 2, 905

Shimomura O., Johnson F. H., Saiga Y., 1962, Journal of Cellular and Comparative Physiology, 59, 223

Sinnecker D., Voigt P., Hellwig N., Schaefer M., 2005, Biochemistry, 44, 7085

Sparks J. S., Schelly R. C., Smith W. L., Davis M. P., Tchernov D., Pieribone V. A., Gruber D. F., 2014, PLoS One 9, e83259

Stewart C. N., 2006, Trends Biotechnol, 24, 155

Taboada, C., et al. 2017, PNAS, p.201701053

Takahashi S., Murata N., 2008, Trends in plant science, 13, 178

Tevini M. (ed.), 1993, In: UV-B Radiation and Ozone Depletion: Effects on Humans, Animals, Plants, Microorganisms, and Materials (Boca Raton, Florida: Lewis Publishers)

Traub W. A., Stier M. T., 1976, Applied Optics, 15, 364

Tsien R. Y., 1998, Annual review of biochemistry, 67, 509

Voet D., Gratzer W. B., Cox R. A., Doty P., 1963, Biopolymers, 1, 193

Wolanin A., Rozanov V. V., Dinter T., Noel S., Vountas M., Burrows J. P., Bracher A., 2015, Remote Sensing of Environment, 166, 243

Xie J., Chen K., Chen X., 2009, Nano research, 2, 261

Zahnle K., Arndt N., Cockell C., Halliday A., Nisbet E., Selsis F., Sleep N. H., 2007, Space Sci Rev, 129, 35

Zahnle K. J., Catling D. C., Claire M. W., 2013, Chemical Geology, 362, 26

Zawada D. G., Mazel C. H., 2014, PLoS ONE, 9, e84570

Zhou X. X., Lin M. Z., 2013, Current opinion in chemical biology, 17, 682